\documentstyle[prc,aps,epsf]{revtex}
\begin{document}

\draft


\title{The H-Dibaryon and the Hard Core}  

\author{D.~E.~Kahana$^{2}$ and S.~H.~Kahana$^{1}$}
 
\address{$^{1}$Physics Department, Brookhaven National Laboratory\\
   Upton, NY 11973, USA\\
   $^{2}$31 Pembrook Dr., Stony Brook, NY 11790, USA}
\date{\today}  
  
\maketitle  
  
\begin{abstract}
The H dibaryon, a single, triply magic bag containing two up, two down and
two strange quarks, has long been sought after in a variety of experiments.
Its creation has been attempted in $K^-$, proton and most recently in
relativistic heavy ion induced reactions. We concentrate on the latter, but our
conclusions are more generally applicable. The two baryons coalescing to
form the single dibaryon, likely $\Lambda \Lambda$ in the case of heavy ions,
must penetrate the short range repulsive barrier which is expected to exist
between  them. We find that this barrier can profoundly affect the probability
of producing the H state, should it actually exist.
\end{abstract}


\section{Introduction}  
Since the H-dibaryon was proposed by R.~Jaffe \cite{Jaffe} as a likely
candidate for a viable six quark bag (uuddss), there have been a variety of
experiments proposed and carried out to locate it, which as yet have
been unsuccessful. Some of these experiments involve production via
[$K^-,K^+$] and proton induced reactions on light targets \cite{hlight}.
More recently strangeness-rich heavy ion collisions \cite{Crawford} have
been thought to offer a preferred mechanism for generating the H. The
H can also be viewed in the limit of ideal $SU_f(3)$ symmetry as the doubly
strange, maximally symmetric, singlet combination of the octet baryons.
However, given the mass gap existing between the $\Lambda$ and
the more massive $\Sigma$ and $\Xi$ states, it is probable that H consists
mainly of $\vert \Lambda \Lambda >$. Consequently its lifetime would be
expected to be closer to one-half that of the $\Lambda$, than to the
$\sim 10^{-9}$ s value suggested by some authors\cite{Donoghue}.

Definitive observation of a double $\Lambda$ hypernucleus is often considered
antithetical to the existence of the H; one reason being that the two strange
baryons, kept captive by their affinity to the normal nucleons would quickly
fall into the lower energy dibaryon state. Of the recorded observations of
such nuclei \cite{Prowse,Danysz,KEK} only the latter, performed at KEK, seems
a good candidate \cite{Dalitz}. This proposition, that the existence
of doubly strange hypernuclei rules out the existence of the H, need not be
valid, and can become prejudicial. A hybrid state consisting partially of
a six-quark bag and partly of a double $\Lambda$ doorway state, attached
to a nucleus might be identifiable in experiments involving the production of
quite light doubly strange hypernuclei \cite{E906}. We will have more to say
about this possibility in what follows.

To our knowledge all theoretical estimates of production rates in heavy ion
collisions \cite{KahanaDover}, irrespective of mechanism, have generally
overlooked the possible existence of a hard core in the baryon-baryon
interaction at short distances. As we will show, under reasonable assumptions
about the hard core, this can lead to quite appreciable suppression of H
production. Here, we introduce this device into the framework of production
via heavy ion collisions. A previous calculation \cite{KahanaDover} suggested
a high formation probability, $\sim 0.07$ H per central Au+Au collision. A
recent AGS experiment, E896 \cite{Crawford}, is presently analyzing some
100 million central $Au+Au$ events and could, in the light of this earlier
prediction of the formation rate, provide a definitive search for the H. It
is our present intention to at least semi-quantitatively understand the
extent the hard core might interfere with this hope.

We treat the short range baryon-baryon force in a transparent and heuristic
fashion. The best evidence from doubly strange nuclei \cite{KEK,Millener}
suggests that the H, if it exists at all, is rather weakly bound, with
binding less than $20$ MeV and probably considerably less. We use this to
justify handling the coalescence of $\Lambda$ pairs in relativistic ion
collisions much in the manner we previously employed for the deuteron
\cite{dekcoal}, also a weakly bound system and likely to form only after all
np-constituents have ceased high energy cascading. This newer calculation of
coalescence, in Ref \cite{dekcoal}, was not employed in earlier work on the H
\cite{KahanaDover}, and in fact had yielded a somewhat reduced formation rate
for the H, by some $50\%$, in the absence of a hard core.  The results
arrived at in the present work suggest a more dramatic suppression. This
conclusion should hold true for either the pure six quark bag proposed by
Jaffe or for hybrid states, in both the relativistic heavy ion and proton
induced environments if the hard core indeed exists.  We refer simply to the
repulsive potential as a hard core, although in practice we employ a
repulsive potential with finite height.

\section{Coalescence and the Hard Core}

Coalescence is treated quantum mechanically in the more recent approach
\cite{dekcoal} by calculating the overlap of the wave packets of the
initial combining pair with an outgoing packet for the final bound
state.  The cascade in which this coalescence estimate is embedded
provides the distributions of both relative momentum and relative
position required for determining the degree of overlap.  The overlap
integral, squared to produce a probability, is then part of a factorised
version of dibaryon production. The combining pair of particles may form
a bound state only after each ceases to interact in the cascade, as was
indicated previously.

However, for two strange baryons to coalesce they must first penetrate
their mutual, repulsive, core. Such a core has a negligible effect on the
deuteron which is spatially rather extended. This cannot be so for the H:
not if this object consists at least partially of six quarks in a bag
comparable in size to that for a single baryon, where short range
repulsion can play a considerable role. H formation from two $\vert \Lambda
\Lambda >$ could be viewed as proceeding in two steps: first a merging into
a broad, deuteron-like doorway state and the second, barrier penetration
into a single compact dibaryon, with the hard core repulsion forming the
barrier. The overall rate for H production is then found to be the product
of the usual coalescence probability and a barrier penetration post-factor. 
Naturally there are unknowns in such a calculation, one being the effective range
at which combining baryons dissolve into a single bag, another being the nature of
the short range forces (the hard core). The first we treat as a parameter;
the second we approximate by using the $NN$ Bonn potential \cite{Bonn},
limiting ourselves to including its shortest ranged components, due
to $\omega$ and $\sigma$ exchange.

Our approach should apply equally well to a hybrid model in which the H 
is a combination of a deuteron-like $\vert \Lambda\Lambda >$ state and a,
presumably smaller, six quark bag. The wave function would then in principle
appear as:

\begin{equation}
\Psi = {\alpha}\vert {\Lambda\Lambda}> + {\beta}\vert q^6 >,
\end{equation}
with $\alpha$ and $\beta$ representing amplitudes for the two-baryon and
six-quark components of the hybrid state. 

Whatever the actual composition of the physical dibaryon, it must have some
minimal six quark bag presence, or else the relatively weak $\Lambda-\Lambda$
force could not lead to binding. The pure Jaffe-like state \cite{Jaffe}
corresponds to $\beta=1$, but our coalescence calculation is in principle
independent of this parameter. The procedure we follow to estimate the effect
of the repulsive core on entry from a doorway $\Lambda-\Lambda$ state into
the final H state would be applicable to either the pure bag or hybrid state
cases. The only question is the precise nature of the hard core itself. We
have stated our assumptions clearly above.

The barrier penetration calculation is described here in full detail. while
the cascade and coalescence formalisms are referenced elsewhere
\cite{dekcoal}. In a standard single meson exchange model (OBEP) \cite{Bonn}
of the nuclear two body interaction the hard core arises from $\omega$
exchange. In transferring this force to the $\Lambda\Lambda$ system
one should, however, probably not scale by the numbers of non-strange
quarks. This component of the force is expected to be essentially
flavour-independent. To the Bonn potential \cite{Bonn} one must then
add a term due to the exchange of a $\phi$ meson between the $s$ quarks.
This observation suggests there should be little or no modification
of the nucleon-nucleon hard core in applying it to the two $\Lambda$
system.  We thus assume an intermediate to short range force exists
of the form:

\begin{equation}
V(r) = V_{\omega}(r)+V_{\sigma}(r),
\end{equation}
where
\begin{equation}
V_i(r)= g_i(1/r)exp(-m_ir). 
\end{equation}
The couplings and the meson masses, $g_i$ and $m_i$ , are specified in
Ref \cite{Bonn}. 

The strong intermediate range $\sigma$ attraction reduces the effect of the
hard core, while the longer range parts of the potential are assumed to play
a negligible role in coalescence. The two baryons will approach to some outer
radius $b$, in fact a classical turning point, before being faced with the
strong short range repulsion produced by the $\omega$. At some smaller radius
$a$, representing the separation of $\Lambda$-cluster centers, the two
baryons are imagined to dissolve into a six quark bag. The calculation is
especially sensitive to this `critical separation radius' $a$. Although our
final results on barrier penetration are consequently somewhat uncertain, it
will become apparent that one thing one cannot do, is to ignore them while
there remains any reason to believe that a short range repulsion exists.

\section{Penetration Factor}  

We appeal to the WKB \cite{Messiah} method for an estimate of the barrier
penetration factor. We require some picture of the inter-baryon potential,
from the larger  separations in the initial doorway state down to the inner
reaches of the final multi-quark bag and for $r \leq a$ take 

\begin{equation}
V(r)= -V_0, 
\end{equation}

while for  $r \geq a$ 

\begin{equation}
V(r)=V_{\omega}(r)+V_{\sigma}(r),
\end{equation}
as specified in Equation 3 above. Thus, inside the radius $a$ the two
baryons, by fiat, melt into a bag. The probability of barrier penetration
in this effective two body model can then be determined by calculating the
WKB approximation for the transmission coefficient at relative energy $E$:

\begin{equation}
T(E)= 4\sqrt{\frac{(V(b)-E)(E-V_0)}{V(b)-V_0}} exp(-2\tau),
\end{equation}

where
\begin{equation}
{\tau} = \int_a^b\,dr\,\sqrt{2m(V(r)-E)}.
\end{equation}
As advertised the upper limit, $b$, of the integral for $\tau$ is a turning
point defined implicitly by
\begin{equation}
E=V(b),
\end{equation}
while $a$ represents the outer separation at which the six quark bag forms. 
The non-relativistic calculation of transmission performed here is probably
adequate, given that coalescence into a relatively weakly bound system
will not proceed at very high baryon-baryon relative momentum.

\section{Coalescence Mechanism}

The rest of the calculation is straightforward, given the existence of a
previously constructed heavy ion two-nucleon coalescence code
\cite{dekcoal,ARC}.  The transmission coefficient $T(E)$ is inserted into
this heavy ion simulation at a point after the formation of a broad doorway
state for the two strange baryons. That is to say, the H formation probability
is taken as the product of $T(E)$ and a coalescence factor for the doorway
state as defined in Reference \cite{dekcoal}. The size and structure of this 
state play only a minor role provided the turning point $b$ is within its
confines. There are two modes for operation of this code, labeled static and
dynamic, both of which are described in Reference \cite{dekcoal}. The dynamic
code is self-contained, providing an internal estimate of the spatial
spreading of the individual baryon wave packets, occasioned by interactions
within the collision medium. The static mode produces essentially identical
results provided the wave packet size, assigned as a fixed value in static
coalescence, is appropriately tuned.

In the earlier work \cite{dekcoal}, a satisfactory agreement of the dynamic
model with known $Si+Au$ deuteron single and double differential
cross-sections was demonstrated, and $Au+Au$ predictions made. Only very
preliminary data for deuterons from AGS $Au+Au$ collisions existed
at that time. In Fig(1) we have compared these 1996 dynamic coalescence
calculations for deuteron production in $Au+Au$ collisions at 11.6 GeV/c,
with very recently submitted data from E866 \cite{chen}. Considering the
nature of both experiment and theory, this prediction of absolute deuteron
yield must be considered as a triumph. There are in the dynamic simulations no
adjustable parameters. In light of this, and to minimise computer time needed
to produce sufficient statistics for the much rarer H dibaryon, we have
performed the coalescence estimates in this work using the static treatment,
adjusted to agree with the dynamic normalisation and perforce with the
deuteron data. This procedure does not appreciably affect H production
estimates and permits us to more efficiently examine bag size parameters $a$
where the hard core suppression can be rather large and the H yield truly
small.

The deuteron prediction gives one great confidence in our treatment of the
coalescence mechanism and lends credence to the use of a similar approach
in estimating the creation of the elusive H. 

\section{ Heavy Ion Production of the H}

We consider $Au+Au$ collisions at an incident energy of $10.6$ GeV per
nucleon. The actual beam energy in E896 \cite{Crawford} is $11.6$ GeV but
the use of a thick target reduces this to a lower effective average. 
The energy dependence, although appreciable and quoted, is by no means
the most critical variable encountered in this simulation. The dependence on
the $\Lambda-\Lambda$ separation, $a$, at the moment of dissolution into a
bag easily wins that title. We have also examined the dependence of the
results on the size $r_h$ of the H-doorway state and $r_{sp}$
the spatial size of the wave packets in the static coalescence model.
Neither prove to be of much consequence. In practice, $r_{sp}$ is chosen
to assure agreement with deuteron yields from dynamic coalescence. In the
present simulations this occurs for $r_{sp}\sim 1.5-2.0 fm$, an
eminently reasonable value. 

The Bonn inspired prescription we described for the $\Lambda \Lambda$
potential reduces numerically to:

\begin{equation}
V(r)=3.94\frac{exp(-3.97r)}{r}-1.44\frac{exp(-2.97r)}{r} GeV,
\end{equation}
with $r$ measured in Fermis. The resulting short distance potential
is graphed in Fig(2).

Our chief results are presented in Fig(3), indicating the variation of the
numbers of H-dibaryons produced in central events with $a$. This latter
parameter must not be thought of as an effective hard core radius for
the $\Lambda-\Lambda$ interaction. The underlying quark-quark forces may also
be viewed as possessing a repulsive short range component due to the exchange
of vector mesons \cite{kahanaripka}. Even with complete overlap of the parent
baryons, i.~e. $a=0$, the average inter-quark separation is, for a uniform
spatial distribution, still comparable to the parent radius $r \sim 0.8$ fm,
i.~e. considerably greater than any conceivable fixed hard core radius. We
have considered baryon centers between $0.2$ fm and $0.5$ fm apart, whereas
reasonable values probably lie between $0.2$ fm and $0.3$ fm, where the
baryon overlap region is near $80$\% of the volume of a single baryon. 
 
Even at the largest separations, H suppression due to the repulsive forces is
not ignorable, but for the smallest values of $a$ the observation of the H,
should it exist, becomes problematic. Early analysis \cite{Judd} of the
actual experimental setup using the simulation ARC \cite{ARC}, suggested a
neutral background comparable with the initial estimate of $0.07$ H's per
central collision \cite{KahanaDover}. Thus, for baryon separations of $0.20$
fm to $0.35$ fm one would need to achieve tracking sensitivities of $10^{-4}$
to $10^{-2}$ relative to background. In our worst case scenario, at $a=0.2$,
one is still left with perhaps a few  thousand  dibaryons produced in
the E896 \cite{Crawford} sample, but immersed in what may prove to be a
daunting background.

We have also considered variations due to bombarding energy
and H-doorway radius. These are easily understood, if not noteworthy.
A decrease of collision momentum from $14.6$ to $10.6$ GeV/c results in
close to a $30\%$ reduction in H's produced per central event, while the
yield is quite insensitive to the doorway radius. 

\section{Conclusions}

Short range repulsion between strange baryons can profoundly hinder  
coalescence into objects whose very existence depends on the presence of
important bag-like structure. This lesson is even more applicable to the
many H-searches initiated using ($K^-,K^+$) reactions \cite{hlight}, since
these generally involve even lower relative energies $E$, and a consequent 
increased difficulty in barrier transmission.

There is perhaps one way to circumvent this frustrating roadblock in the
discovery of the lightest of all possible strangelets. In the event a pair of
strange dibaryons are attached, through a [$K^-,K^+$] reaction, to a light
nucleus, a hybrid H may form, and itself remain bound to the nucleus. An
optimum final nucleus might be $^5_{\Lambda\Lambda}H$ \cite{E906,LLH5}. The
extra nucleons in this five particle system keep the captured $\Lambda$ pair
together for some $100$ picoseconds, this being far more than enough time
for penetration of what then would constitute a rather modest, $1-2$ GeV
barrier. Any evidence that the $\Lambda\Lambda$ pairing energy substantially
exceeded the $2$--$3$ MeV or so expected from the known $\Lambda\Lambda$
interaction \cite{nimjegen} would indicate the presence of a hybrid
bag+doorway state. Appreciable observed decay into the $\Sigma^-p$ channel
would strengthen such a conclusion.

Unfortunately, the very same repulsive forces which made coalescence into a
bound H state difficult, may also, at the quark level, destroy the existence of
the state. Only a detailed microscopic calculation can begin to answer this
question, for example in a model in which quarks rather than baryons exchange
mesons \cite{kahanaripka}. It is, in this context, disturbing that the search
for the only `strangelets' which we are certain do exist, i.~e. doubly strange
hypernuclei, may be discontinued despite the present finding of only a single
good candidate \cite{KEK}.

\section{Acknowledgements}

The authors are grateful to C.~Chasman for providing us with the  recently submitted E866
deuteron data. The present manuscript has been authored under US DOE grants NO. DE-FG02-93ER407688 
and DE-AC02-76CH00016.

\begin{figure}
\vbox{\hbox to\hsize{\hfil
\epsfxsize=6.1truein\epsffile[24 85 577 736]{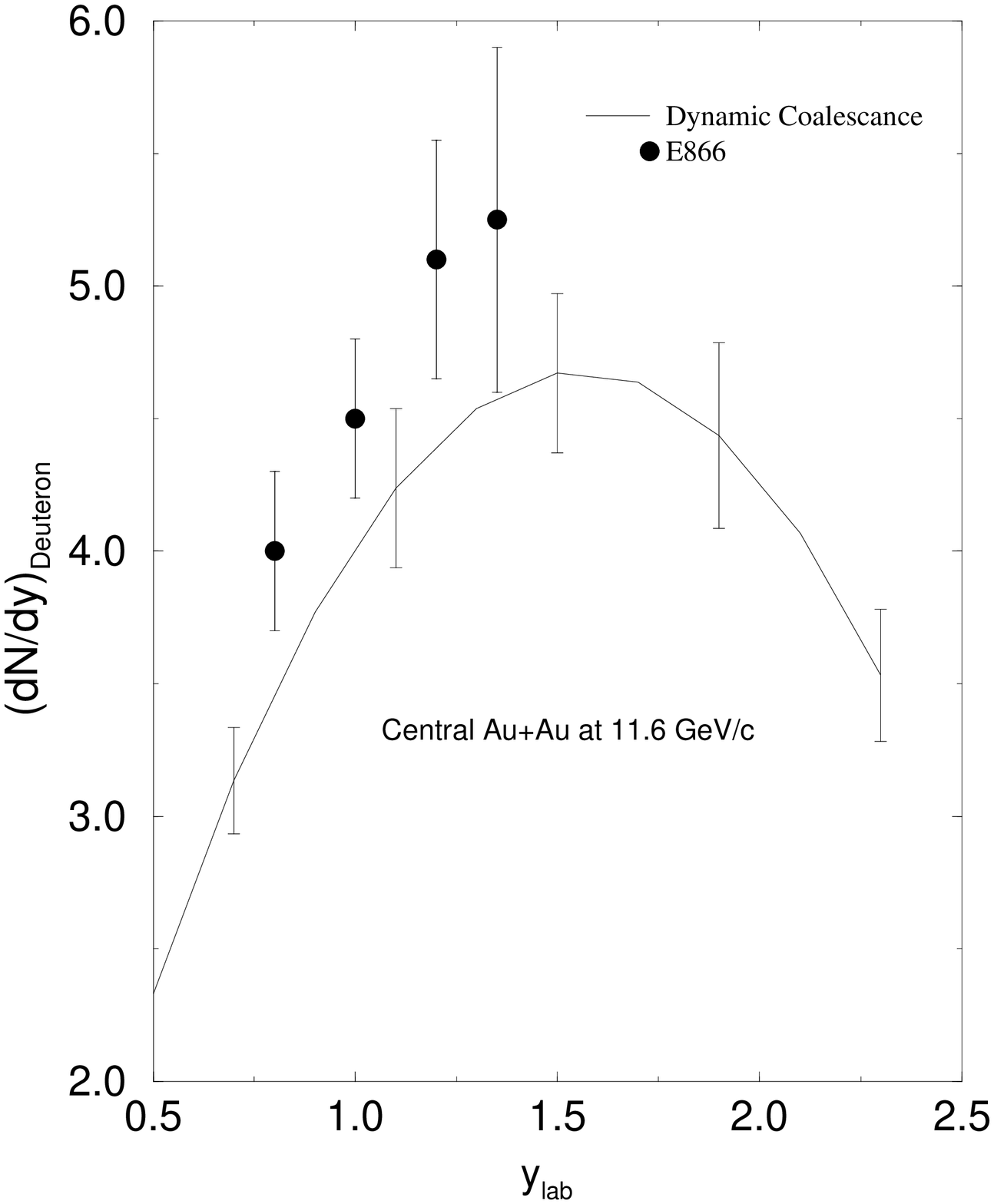}
\hfil}}
\vspace*{10mm}
\caption[]{Comparison of E866 data \cite{chen}, recently submitted to
Phys.~Rev.~C. (1999), with absolute predictions of the Dynamic Coalescence
model from 1996 \cite{dekcoal}. The agreement is excellent considering the
conditions of both theory and experiment. A future work will examine the
entire data set, including an attempt to insure that the experimental
definition of centrality is more explicitly built into the theoretical
simulation. For present purposes the $5-10\%$ difference in normalisation
between data and calculation is of no consequence.}
\label{fig:one}
\end{figure}
\clearpage

\begin{figure}
\vbox{\hbox to\hsize{\hfil
\epsfxsize=6.4truein\epsffile[24 85 577 736]{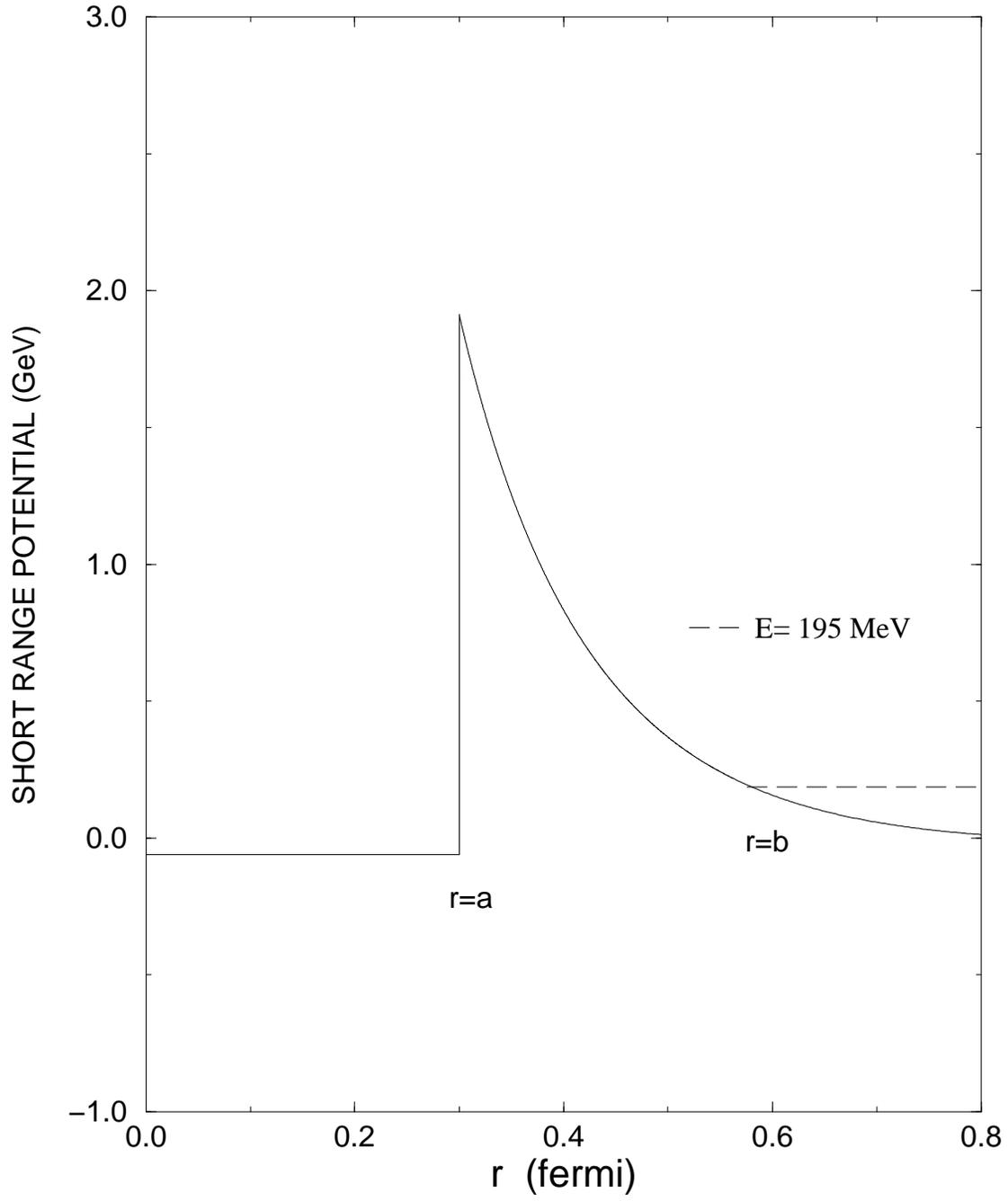}
\hfil}}
\caption[]{The short range $\Lambda$-$\Lambda$ potential taken from 
the $\sigma$ and $\omega$ exchange parts of the Bonn potential. At some 
separation $a$ the two strange baryons are assumed to dissolve into
a single bag, represented here by a shallow, attractive, effective
potential. Barrier penetration from the di-$\Lambda$ doorway state is 
represented by the dashed line.}
\vspace*{10mm}
\label{fig:two}
\end{figure}
\clearpage

\begin{figure}
\vbox{\hbox to\hsize{\hfil
\epsfxsize=6.4truein\epsffile[24 85 577 736]{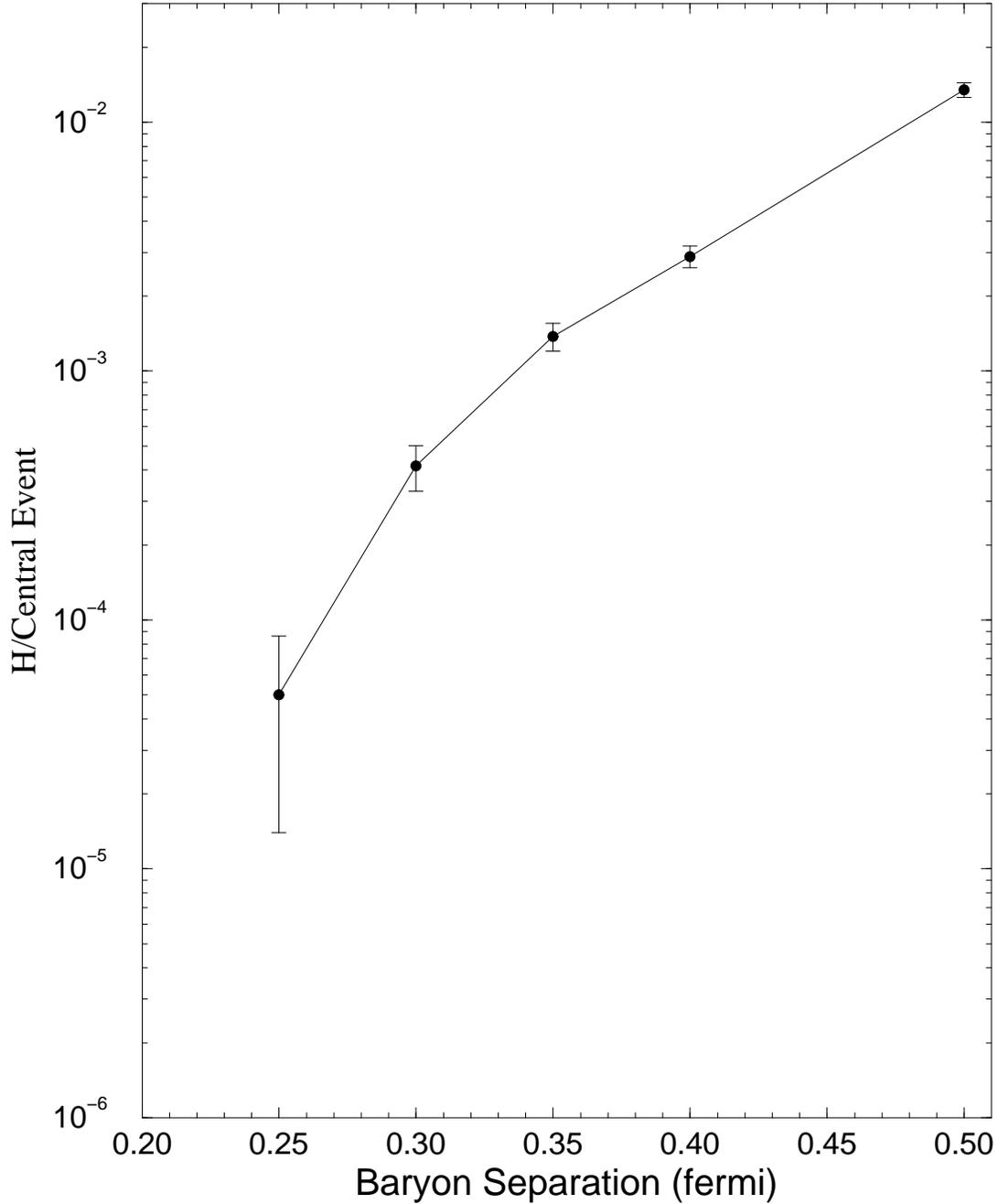}
\hfil}}
\caption[]{Absolute H Production per Central $Au+Au$ event. The precise
separation of di-$\Lambda$ centers, $a$, at which a single six quark bag
forms is of course not known, but a reasonable value is likely less than
$0.3$  fm. Even for complete baryon overlap, the average distance between
constituent quarks is greater than the conventional nucleon-nucleon hard core
radius of $0.4$ fm. The assumed energy for the collision is an average
$10.6$ GeV to account for averaging by a thick target. The energy dependence
is not strong enough to alter the displayed results appreciably. Centrality
is defined by $b\leq 2$, but again calculation indicates only a weak dependence
on impact parameter.}
\label{fig:three}
\end{figure}

\begin{figure}
\vbox{\hbox to\hsize{\hfil
\epsfxsize=6.4truein\epsffile[24 85 577 736]{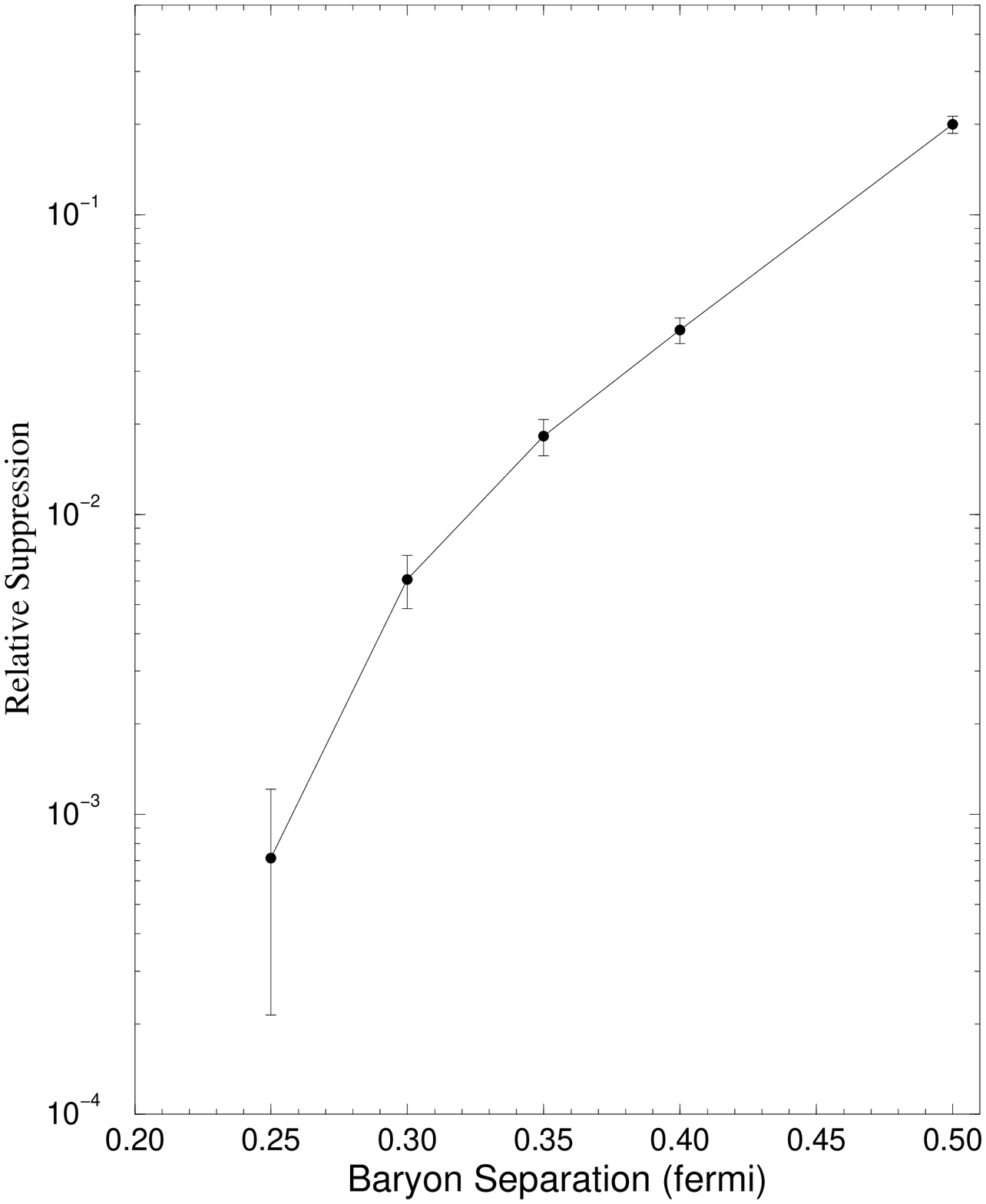}
\hfil}}
\caption[]{ H Suppression. Suppression is here taken relative to the 
older calculation of $0.07$ per central event \cite{KahanaDover} since 
this gives one an approximate suppression relative to an early estimate 
of background \cite{Judd}. Again the abscissa is the assumed separation 
$a$ at which a bag forms.}
\label{fig:four}
\end{figure}

\clearpage


\begin{references}
\bibitem{E906}
A.~Rusek, Private Communication. The BNL experiment E906, still under 
analysis, might observe nuclei such as $^5_{\Lambda\Lambda}H$.
\end{references}
\end{document}